\documentclass[twocolumn]{aastex631}
\usepackage[utf8]{inputenc}
\usepackage{amsmath}
\usepackage{amssymb}
\usepackage{natbib}
\usepackage{mathtools}

\usepackage{graphicx}
\usepackage{comment}
\usepackage{txfonts}
\usepackage{ctable}
\usepackage{xspace}
\usepackage{epstopdf, epsfig}
\usepackage{tabularx}
\usepackage{threeparttable}
\usepackage{tablefootnote}
\usepackage{xcolor}

\newcommand{\mponn}{\textcolor{black}}
\newcommand{\mpon}{\textcolor{black}}
\newcommand{\mpo}{\textcolor{black}}
\newcommand{\rb}{\textcolor{black}}
\newcommand{\is}{\textcolor{black}}

\newcommand{\IS}{\textcolor{black}}

\newcommand{\RBn}{\textcolor{black}}

\newcommand{\isref}{}

\def\aap{A\&A}%
\def\apjs{ApJS}%
\def\apjl{ApJ}%
\shorttitle{Supernova remnants in circumstellar magnetic fields}
\shortauthors{Sushch et al.}

\begin{document}

\title{Leptonic non-thermal emission from supernova remnants evolving in the circumstellar magnetic field}

\correspondingauthor{Iurii Sushch}
\email{iurii.sushch@nwu.ac.za}

\author[0000-0002-2814-1257]{Iurii Sushch}
\affiliation{Centre for Space Research, North-West University, 2520 Potcheftroom, South Africa}
\affiliation{Astronomical Observatory of Ivan Franko National University of Lviv, Kyryla i Methodia 8, 79005 Lviv, Ukraine}
\affiliation{Deutsches Elektronen-Synchrotron DESY, Platanenallee 6, 15738 Zeuthen, Germany}

\author[0000-0002-8312-6930]{Robert Brose}
\affiliation{Dublin Institute for Advanced Studies, Astronomy \& Astrophysics Section, 31 Fitzwilliam Place, D02 XF86 Dublin 2, Ireland}
\affiliation{University of Potsdam, Institute of Physics and Astronomy, 14476 Potsdam, Germany}


\author[0000-0001-7861-1707]{Martin Pohl}
\affiliation{Deutsches Elektronen-Synchrotron DESY, Platanenallee 6, 15738 Zeuthen, Germany}
\affiliation{University of Potsdam, Institute of Physics and Astronomy, 14476 Potsdam, Germany}

\author[0000-0001-6975-5186]{Pavlo Plotko}
\affiliation{Deutsches Elektronen-Synchrotron DESY, Platanenallee 6, 15738 Zeuthen, Germany}

\author{Samata Das}
\affiliation{Deutsches Elektronen-Synchrotron DESY, Platanenallee 6, 15738 Zeuthen, Germany}
\affiliation{University of Potsdam, Institute of Physics and Astronomy, 14476 Potsdam, Germany}



\begin{abstract}

The very-high-energy \is{(VHE; $E>100$~GeV)} gamma-ray emission observed from a number of \mpo{Supernova remnants (SNRs)} \mpon{indicates particle acceleration to high energies at the shock of the remnants and a potentially significant contribution to Galactic cosmic rays. It is extremely difficult to determine whether protons (through hadronic interactions and subsequent pion decay) or electrons (through inverse Compton scattering on ambient photon fields) are responsible for this emission.} \IS{
For a successful diagnostic, a good understanding of the spatial and energy distribution of the underlying particle population is crucial. Most SNRs are created in core-collapse explosions \mpon{and expand into the wind bubble of} their progenitor stars. This circumstellar medium features a complex spatial distribution of gas and magnetic field which naturally strongly affects the resulting particle population. In this work, we conduct a detailed study of the \mpon{spectro-spatial evolution of the electrons accelerated at the forward shock of core-collapse SNRs and their non-thermal radiation, using the RATPaC code that is designed for the time- and spatially dependent treatment of particle acceleration at SNR shocks.} We focus on the \mpon{impact of the spatially inhomogeneous magnetic field through the efficiency of} diffusion and synchrotron cooling. \mpon{It is demonstrated that the structure of} the circumstellar magnetic field can \mpon{leave strong signatures in} the spectrum and morphology of the resulting non-thermal emission.}

\end{abstract}

\keywords{Supernova remnants (1667) --- Stellar wind bubbles (1635) --- Red supergiant stars (1375) --- Wolf-Rayet stars (1806) --- Gamma-rays (637) --- Galactic cosmic rays (567)}


\section{Introduction} \label{sec:intro}

\mpo{The very-high-energy gamma-ray emission, that has been observed from \is{a handful of} Supernova remnants (SNRs),} can be explained by both electrons (through inverse Compton (IC) scattering on ambient photon fields) and protons (through hadronic interactions with subsequent decay of neutral pions) and it is very difficult to discriminate between these two scenarios. Although acceleration of electrons in SNRs can be directly confirmed through the detection of synchrotron emission in the X-ray energy band, similar confirmation for protons would \IS{need} a careful analysis of the observed gamma-ray emission. 
One of the key \IS{signatures of the hadronic radiation process} is the so-called pion bump between $100$ MeV and a few GeV 
\citep{1973ApJ...185..499S}. The exact shape of the pion bump and the energy of its peak is dependent on the energy distribution of the parent particle population \citep{2018A&A...615A.108Y} and \mpon{its elemental composition \citep{2020APh...12302490B},} but in any case at low energies it is marked by the abrupt decrease of the gamma-ray flux \mpo{that is not expected in the spectra of inverse Compton scattering.} The first reliable detection of such a low-energy cut-off associated with a pion bump, \mpo{and hence direct evidence for proton acceleration,} was found for two old SNRs, IC 443 and W44, which are interacting with molecular clouds \citep{paper:w44_ic443_pi0_fermi}. \is{Another discriminating feature is the shape of the gamma-ray spectrum above the pion bump energies. The hadronic \mpo{gamma-ray} spectrum above the pion bump is expected to reflect the spectrum of the parent particle population \IS{with the spectral index of photons, $\alpha$, similar to that of the radiating particles, $s$.}
In contrast, \mpon{the spectral index of the IC gamma-ray emission is with $\alpha = (s+1)/2$ generally softer than that of the radiating electrons.}
Recently, a deep Fermi-LAT survey revealed a large population of gamma-ray bright SNRs \citep{2016ApJS..224....8A}, a considerable portion of which were identified as dynamically old \mpo{SNRs and predominantly hadronic emitters through their spectral shape and potential association} with molecular clouds \citep{2016ApJ...816..100J, 2019A&A...623A..86A, 2020MNRAS.497.3581D}}

Typically, the particle spectrum (both for electrons and for protons) follows a power law with spectral index around $s=2$, as predicted by diffusive shock acceleration (DSA) {\isref in the test-particle approximation \citep{1983SSRv...36...57D}}, with an exponential cut-off at some maximum energy which is imposed by the age of the remnant, particle escape, or synchrotron cooling (relevant only for electrons). {\isref Deviations from the universal value of $s=2$ may arise from accounting for e.g. cosmic-ray feedback \citep{2001RPPh...64..429M,2006MNRAS.371.1251A}, \mponn{secondary acceleration processes \citep{2015A&A...574A..43P,2020A&A...639A.124W}, fast motion of downstream turbulence} \citep{2020ApJ...905....1H,2020ApJ...905....2C}, \mponn{and inefficient particle confinement \citep{2011NatCo...2..194M}. With time the spectrum of confined particles} considerably softens to $s\approx2.7$ due to the escape of high-energy CRs and a rapid reduction of the maximally achievable energy  \citep{2020A&A...634A..59B, 2019MNRAS.490.4317C}, as well as, in the case of electrons, continuous synchrotron losses \citep{2004A&A...427..525B, 2019PhRvL.123g1101D,2020A&A...634A..59B}}. 

\mpon{Substantial variation of this picture can arise from }
the complexity of its environment. 
The majority of SNRs in the Galaxy are produced in core-collapse events \is{\citep{1991ARA&A..29..363V}}, \is{typically} either at the red supergiant (RSG) or at the Wolf-Rayet (WR) stage of stellar evolution \is{\citep{2009ARA&A..47...63S}}.
In such a case the remnant evolves inside a stellar wind bubble \IS{of complicated structure. In particular, the complexity of the circumstellar magnetic field can substantially modify the electron distribution through energy losses and changes of the particle confinement, \mpon{leaving characteristic imprints in the spectrum and morphology of the observed radiation and their} temporal evolution. The resulting IC gamma-ray spectrum 
would strongly depend on the age of the remnant and could considerably differ from \mpon{that for simple DSA scenarios and in fact be similar to that} of hadronic emission, \mpo{which would} add an extra dimension to the discrimination problem.}

We study the impact of the circumstellar magnetic field on the \mpon{distribution of electrons accelerated at the forward shock and their non-thermal emission. For that purpose we use the RATPaC code (Radiation Acceleration Transport Parallel Code) \citep{2012APh....35..300T, 2012A&A...541A.153T, 2013A&A...552A.102T, 2018A&A...618A.155S,2019A&A...627A.166B} to simulate the SNR and the acceleration and transport of energetic particles in it.}

\mpon{Earlier studies of cosmic-ray production in core-collapse SNR \citep{2012A&A...541A.153T, 2013A&A...552A.102T} were based on a relatively weak magnetic field in the wind zone, and hence insignificant synchrotron losses, and \IS{suffered from an overestimation of the effects of }
interactions of the forward shock with the contact discontinuity or other shocks \IS{due to not ideal temporal resolution.}} \is{More recently, \citet{2018MNRAS.475.5237G} studied the time evolution of the gamma-ray spectrum from the Type II SNR (RSG progenitor) with strong simplifications. \mpon{At each time step accelerated particles were injected \mpo{with a power-law spectrum up to} a certain maximum energy that was subsequently evolved accounting for adiabatic and radiative losses. They noticed some spectral features in the leptonic non-thermal emission}, but did not discussed them further, as the study was focused on the hadronic emission and the effectiveness/duration of the "PeVatron" phase of the SNR.}

\is{Below we revisit this problem taking advantage of the fine temporal and spatial resolution \mpon{offered by the RATPaC code} and solve the transport equation for particles simultaneously with the hydrodynamic equations.}

\section{Circumstellar environment of core-collapse SNRs}
\label{sec:CM}

Core-collapse SNRs are results of explosions of massive stars at the end of their evolution. Throughout their life these stars \mpon{drive} a stellar wind that shapes a circumstellar bubble around the star. 
The hydrodynamical evolution of these bubbles \mpon{has been studied} both analytically and numerically \citep[e.g.][]{1977ApJ...218..377W, 1995ApJ...455..145G, 1996A&A...316..133G, 1996A&A...305..229G, 2005ApJ...630..892D, 2007ApJ...667..226D, 2007ASSP....1..183A}. The general bubble \mpon{created during a single evolutionary phase of the star} can be roughly divided into four zones. Closest to the star is the \IS{free wind (FW)}. At the termination shock the wind material is \mpo{heated} and feeds the hot shocked wind \IS{(SW)}. The \IS{SW} zone \mpon{ends at a contact discontinuity separating the shocked material of the ISM or an earlier wind phase. A bit further out one finds a shock in the outer medium and eventually reaches the} unperturbed ISM. The density in the FW zone falls off $\propto r^{-2}$, and in the \IS{SW} zone it is approximately constant. This general picture is complicated by changes in the wind properties on account of stellar evolution \citep{2020MNRAS.493.3548M}. \mpon{The last stage of stellar evolution, RSG or WR, is decisive for the shape of} the bubble prior to the supernova explosion.

\is{The RSG wind is \mpon{much slower than that of the earlier main-sequence phase and does not provide a simple Weaver-type bubble.} However, the deceleration of the main-sequence wind and the eventual backflow of the \mpon{gas still result in the formation of a shock and a thin shell of RSG and main-sequence wind material} \citep{2007ApJ...667..226D}. At the end of the RSG phase, the FW zone of the RSG wind terminates at the thin shell, beyond which there is the shocked main-sequence wind. In our simulations we ignore the thin shell and approximate the hydrodynamic profile by an abrupt drop in density at the transition point from the free RSG wind to the shocked main-sequence wind.} The mass-loss rate during the RSG phase ranges from $10^{-7}~M_{\odot}/\rm{yr}$ to $10^{-5}~M_{\odot}/\rm{yr}$  or more \citep{2018MNRAS.475...55B}). The wind speed is low, $15$ to $50$~km$/$s \citep{2011A&A...526A.156M}. 

For WR progenitors the environment is shaped by a fast wind that sweeps up the preceding RSG wind. The wind speed ranges from $1000$ to $4000$ km$/$s, and the mass-loss rate is similar to that of RSGs, $5\times10^{-6}~M_{\odot}/\rm{yr}$ to $5\times10^{-5}~M_{\odot}/\rm{yr}$ \citep{2007ARA&A..45..177C}. \mpon{At the termination shock, the gas density increases toward the \IS{SW}.}

\mpon{The 
changes in the density affect the injection rate into DSA and modify the shock speed, leading to variations} in the maximum energy to which particles can be accelerated. The temperature of the \IS{SW} is much higher \mpon{than that of the FW, and so is the speed of sound. The consequences for the compression ratio and the spectral index of accelerated particles \citep[e.g.][]{Das:20216k}} will be examined in detail in forthcoming studies. In this paper we focus on the impact of the magnetic field.

We construct two generic hydrodynamic profiles describing the circumstellar medium for RSG and WR progenitors. \is{Recent numeric simulations and observations suggest that the X-ray and VHE gamma-ray \mpo{brightness of leptonic emission} should strongly decrease after $1000-4000$ years due to a rapidly  decreasing maximum electron energy and particle escape \citep{2018A&A...612A...3H,2020A&A...634A..59B}.} \mpon{Limiting our study to young SNRs, we focus on} the FW zone and the transition to the \mpo{next zone, \is{which basically constitutes the \IS{SW} of all the previous phases}}. 
\mpo{The location of \is{the transition point to the \IS{SW} zone}} is typically around {\isref $\mathbf{5-15}$ pc \citep[e.g.][]{2007ASSP....1..183A, 2007ApJ...667..226D}} \mpon{and here firmly set to} $r_{\mathrm{TP}}=5$~pc for both cases, \is{to shorten the simulation time}. 
\mpon{In the FW the density scales as $\propto r^{-2}$ with a mass-loss rate $\dot{M}_\mathrm{RSG} = 10^{-5}~M_{\odot}/\rm{yr}$ and wind speed $V_\mathrm{RSG} = 20$~km$/$s for the RSG. For the WR star we set $\dot{M}_\mathrm{WR} = 10^{-5}~M_{\odot}/\rm{yr}$ and $V_\mathrm{WR} = 2000$~km$/$s. }
The temperature \is{in the FW} is set to $10^4$~K, while the pressure follows \mpon{from the ideal gas law}. 
For the RSG progenitor, we assume a sharp decrease of the density by a factor of 100 at $r_{\mathrm {TP}}$, \mpon{and the flow speed falls to zero}. 
Therefore the circumstellar medium of the RSG can be described by the following
\begin{align}
    \rho(r) &= \begin{cases}
            {\frac{\dot{M_\star}}{4\pi r^2 v_{wind}}}, & r\leq r_{\mathrm{TP}},\\
            \rho(r_{\mathrm{TP}})\times10^{-2}, & r>r_{\mathrm{TP}},           
            \end{cases} \label{RSG_rho} \\
    v(r) &= \begin{cases}
            V_\mathrm{RSG}, &r\leq r_{\mathrm{TP}},\\
            0, & r>r_{\mathrm{TP}},
            \end{cases}\\
    p(r) &= \rho(r) RT(r)\\
    T(r) &= 10^4 K \label{RSG_temp}
\end{align}

\mpon{For the WR progenitor we have a termination shock with the usual jump conditions for a strong hydrodynamic shock. The flow speed downstream of the shock decreases as $\propto r^{-2}$ to satisfy the continuity of the mass flux. In total, the circumstellar medium of a WR progenitor is described by}

\begin{align}
    \rho(r) &= \begin{cases}
            {\frac{\dot{M_\star}}{4\pi r^2 v_{wind}}}, & r\leq r_{\mathrm{TP}},\\
            4\rho(r_{\mathrm{TP}}), & r>r_{\mathrm{TP}},           
            \end{cases} \label{WR_rho} \\
    v(r) &= \begin{cases}
            V_\mathrm{WR}, &r\leq r_{\mathrm{TP}},\\
            0.25 V_\mathrm{WR} (r_{\mathrm{TP}}/r)^2, & r>r_{\mathrm{TP}},
            \end{cases}\\
    p(r) &= \begin{cases}
            \rho(r) RT(r), &r\leq r_{\mathrm{TP}},\\
            3/4 \rho(r_{\mathrm{TP}})V_\mathrm{WR}^2, & r>r_{\mathrm{TP}},
            \end{cases}\\
    T(r) &= \begin{cases}
            10^4 K, &r\leq r_{\mathrm{TP}},\\
            p(r)/R\rho(r),  & r>r_{\mathrm{TP}}.
            \end{cases} \label{WR_temp}
\end{align}

\subsection{Magnetic field}
\label{sec:Bfield}

Recently, \citet{2015A&A...584A..49V} performed magneto-hydrodynamical simulations of the wind bubble around the massive star, \mpon{but ignored the magnetic field in the wind material}. Both RSGs and WR stars exhibit strong surface magnetic fields which should be transported with the stellar wind. \rb{Additionally, a turbulent dynamo might convert a sizable fraction of the \mpo{bulk flow} energy to magnetic field \cite[and references therein]{2018MNRAS.479.4470M}.} 

In the free-wind zone the magnetic field \mpon{is mostly azimuthal due to the stellar rotation. Its amplitude is} well described by \citep{1982ApJ...253..188V}
\begin{equation}
    B_\mathrm{wind}(r) = B_\ast\,\frac{R_\ast}{r^2} \,\sqrt{
    \frac{V_\mathrm{rot}^2 r^2+R_\ast^2 V_\mathrm{wind}^2}{V_\mathrm{rot}^2 + V_\mathrm{wind}^2}} \ ,
\end{equation}
with $B_\ast$ denoting magnetic field strength at the surface of the star, $R_\ast$ and $V_\mathrm{rot}$ are the radius and the rotation velocity of the star, and $V_\mathrm{wind}$ is the wind speed. 
We ignore the very inner region with predominantly radial magnetic field, because that is relevant only during the first days and weeks of the SNR. It is still important to have a rough estimate of the $V_\mathrm{wind}/V_\mathrm{rot}$ ratio to correctly scale the magnetic field strength.

The initial rotation velocities of massive stars are quite high, reaching few hundreds km$/$s \citep{2008A&A...479..541H,2012RvMP...84...25M}. \mpo{At the RSG phase, the large size and high mass-loss rate makes the stars} slow rotators with typical velocities, predicted by all stellar models, of a few km$/$s \citep[][and references therein]{2012RvMP...84...25M}. These estimates lead to 
\begin{equation}
    B_\mathrm{wind}(r) = (0.1 - 0.5) B_\ast \frac{R_\ast}{r}\ ,\  \mathrm{for}\quad r\gtrsim R_\ast V_\mathrm{wind}/V_\mathrm{rot} 
\end{equation}
for the FW of a RSG.

The rotation velocity of WR stars is very uncertain \citep[see e.g.][]{2007ARA&A..45..177C}. \mpon{For some stars the co-rotation of large-scale structures indicates values ranging from few tens to few hundreds km$/$s \citep{2008IAUS..250..139C, 2010ApJ...716..929C, 2011ApJ...735...13H}, as do strong deviation from spherical symmetry \citep{1998MNRAS.296.1072H}. On that basis we find}
\begin{equation}
    B_\mathrm{wind}(r) \sim (0.01 - 0.1) B_\ast \frac{R_\ast}{r} 
\end{equation}

RSGs exhibit relatively \mpo{weak} surface magnetic field of $1-10$~G \citep{2017A&A...603A.129T}, but a very large size of \mpo{a few hundred $R_\odot$}. On the contrary, Wolf-Rayet stars are argued to be strongly magnetic, up to $1000$~G, because the final product of their evolution are highly magnetized pulsars or magnetars \citep{2007ARA&A..45..177C}. There is some evidence that magnetic field in some parts of the wind can reach $100$~G implying a surface magnetic field of $1000$~G \citep{2014ApJ...781...73D}. WR stars are also quite compact with a radius of only $1-10~R_{\odot}$ \citep{2007ARA&A..45..177C}.

At the transition to the \IS{SW} \is{zone}
we assume a step-like change of the magnetic field. \mpon{For the WR progenitor the field strength increases by a factor of 4 at the termination shock and will then linearly increase with radius as the wind speed falls $\propto r^{-2}$.
In the RSG scenario we have a transition to the main-sequence wind with highly uncertain magnetic-field strength. In line with the assumption of $B^2 \propto \rho$ we assume a decrease by a factor of 10.} \is{Beyond the transition point we assume that the magnetic field stays constant as there is no flow.}

Denoting as $B_0$ the initial magnetic field close to the star, the circumstellar magnetic field is expressed as 
\begin{equation}
\label{bwindRSG}
    B_{\rm CM}(r) = 
    \begin{dcases}
        B_0 \frac{R_\ast}{r}, & r\leq r_{\mathrm{TP}},\\
     \mpo{   0.1 B_0 \frac{R_\ast}{r_{\mathrm{TP}}}}, & r > r_{\mathrm{TP}}
    \end{dcases}
\end{equation}
for the RSG and 
\begin{equation}
\label{bwindWR}
    B_{\rm CM}(r) = 
    \begin{dcases}
        B_0 \frac{R_\ast}{r}, & r\leq r_{\mathrm{TP}},\\[5pt]
        4 B_0 \frac{R_\ast}{r_{\mathrm{TP}}} \is{\frac{r}{r_{\mathrm{TP}}}}, & r > r_{\mathrm{TP}}
    \end{dcases}
\end{equation}
for the WR star. 
In the following we consider three configurations of the adopted values for $B_0$ and $R_\ast$ for both scenarios, RSG and WR (see~Table~\ref{tab:stellar_params}). HNA and MNA models correspond respectively to a high and a moderate magnetic field strength within the range of allowed values discussed above. The AMP model adopts a \mpo{field strength lying between those of the} HNA and MNA models but assumes further amplification of the field upstream of the shock (see Section~\ref{sec:Bfield_model}).

 \begin{table}
    
      \centering
      \caption{Input parameters describing the circumstellar magnetic field}
      \begin{tabular}{l|ccc|ccc}
        \hline
        \hline
        & \multicolumn{3}{c|}{RSG} & \multicolumn{3}{c}{WR} \\
        Parameter& HNA$^a$ & MNA$^b$ & AMP$^c$ & HNA & MNA & AMP \\
        
        \hline
        $B_0$ [G] & 7& 1& 3& 100 & 10 & 50 \\
        $R_\ast$ [$R_\odot$] & 1000 & 500 & 1000 & 10 & 5 & 10\\ 
        \hline
        \\
      \end{tabular}
      \label{tab:stellar_params}
    
    \raggedright
    \footnotesize{$^a$ High non-amplified circumstellar magnetic field}
    
    \footnotesize{$^b$ Moderate non-amplified circumstellar magnetic field}
    
    \footnotesize{$^a$ Circumstellar magnetic field amplified upstream with an amplification factor $k=5$}
    \end{table}

\section{Modelling \is{of the SNR evolution}}

\subsection{Hydrodynamics}

{The standard gasdynamical equations}
are solved, \mpon{meaning we} 
assume that the magnetic field is dynamically unimportant due to its low strength and that the remnant is not in the radiative phase yet \citep{2016MNRAS.456.2343P}. The equations are solved in 1D for spherical symmetry.

{We initialize the ejecta profile by a \mpo{constant density,}  $\rho_{\mathrm c}$, up to the radius $r_{\mathrm c}$, followed by a power-law distribution up to the ejecta-radius $R_{\mathrm{ej}}$ \mpon{with index $n=9$ for the core-collapse explosion},
\begin{align}
 \rho(r) &= \begin{dcases}
             \rho_{\mathrm c}, & r<r_{\mathrm c},\\
             \rho_{\mathrm c}\left(\frac{r}{r_{\mathrm c}}\right)^{-n}, & r_{\mathrm c}\leq r \leq R_{\mathrm{ej}},\\
            \end{dcases}
	\label{gasdyn}
\end{align}}
\is{beyond which the density profile is \mpo{that of} the circumstellar medium described in Section~\ref{sec:CM}}. The velocity of the ejecta is defined as
\begin{equation}
v_{\mathrm{ej}}(r) = \frac{r}{T_{\mathrm{SN}}},
\end{equation}
where $T_{\mathrm{SN}}=1$\,yr is the initial time set for hydrodynamic simulations. Then for the assumed \mpo{ejecta mass, $M_{\mathrm{ej}}=3\,M_\odot$,
and energy, $E_{\mathrm{ej}}=10^{51}$~erg (for \is{both RSG and WR}),} and defining the radius of the ejecta as multiple of $r_{\mathrm c}$, $R_{\mathrm{ej}} = xr_{\mathrm c}$, the initial conditions for simulations can be written as
\begin{align}
  r_c &=
  \left[\frac{10}{3}\frac{E_{\mathrm{ej}}}{M_{\mathrm{ej}}} \left(\frac{n-5}{n-3}\right) \left(\frac{1- \frac{3}{n} x^{3-n}}{1-\frac{5}{n} x^{5-n}} \right) \right]^{1/2} T_{\mathrm{SN}}, \\
  \rho_c &= \frac{M_{\mathrm{ej}}}{4\pi r_c^3}\frac{3(n-3)}{n} {\left(1- \frac{3}{n} x^{3-n}\right)^{-1}}, \\
  v_{\mathrm c} &= \frac{r_{\mathrm{c}}}{T_{\mathrm{SN}}}. 
\end{align}
\is{The spatial grid used in the hydrodynamic simulations extends to $r=30\,$pc with $262144$ linearly spaced grid points.}

\subsection{Magnetic field}
\label{sec:Bfield_model}

\mpon{Upstream of the forward shock of the SNR the ambient magnetic field is assumed to be amplified by a factor $k$ and to \mpo{exponentially decline on a length scale $\Delta l = 0.05 R_\mathrm{sh}$, until it reaches the field strength} in the far-upstream medium.} \mpo{Close to the transition point, $r_\mathrm{TP}$, at which the ambient field abruptly changes, the thickness} of the precursor region is redefined as $\Delta l = (r_\mathrm{TP} - R_\mathrm{sh})$. After the transition point it is changed back to 5\% of the shock radius.

\mpon{Assuming a shock compression ratio of 4, the} immediate downstream magnetic field is then given by 
\begin{equation}
    B_{\rm d} = \sqrt{11} k B_{\rm wind}(R_\mathrm{sh})
\end{equation}
The downstream magnetic field is passively transported with the plasma flow, following the induction equation for ideal MHD \citep{2013A&A...552A.102T}. \is{For HNA and MNA models we assume that the magnetic field is not amplified upstream of the shock ($k=1$), while for the AMP model the amplification factor is $k=5$. }
\mpon{The field amplification factor, $k$, has a particular impact on the rate of synchrotron cooling. }

{\isref
\RBn{Describing the amplification of the magnetic field only by the parameter $k$ is of course a crude simplification of the microphysics involved. As far as the resonant streaming instability is concerned, the amplification of turbulence should saturate a level of $\delta B\approx B_0$ \citep{Zweibel:1979zw}. An amplification to $\delta B\approx \sqrt{v_\mathrm{sh}U_\mathrm{cr}/c} \geq B_0$ in the precursor is possible, if the non-resonant streaming instability is involved \citep{2000MNRAS.314...65L, 2001MNRAS.321..433B}.} \mponn{\citet{2021A&A...650A..62C} argue that the saturation level ought to be $\propto v_{\mathrm{sh}}^{3/2}$ for the Bell scenario and by $\propto v_{\mathrm{sh}}$ for the resonant streaming instability.}

\RBn{The particulars of the nonresonant amplification are clearly beyond the scope of this paper. The nonresonant instability saturates by a back-reaction of the thermal plasma to CR-streaming or a modification of the bulk flow \citep{2009ApJ...694..626R, 2010ApJ...709.1148N,2017MNRAS.469.4985K}. In both cases, the saturation level is independent of the initial level of turbulence or the ambient magnetic field. However, the growth of turbulence is limited by the time that is available for the growth of turbulence, since the turbulence needs to be replenished on timescales of $D(p)/v^2_\text{sh}$ \citep{2016A&A...593A..20B}. \mponn{For Bell modes at the saturation amplitude $\sqrt{v_\mathrm{sh}U_\mathrm{cr}/c} $ throughout the cosmic-ray precursor one would not even get a single exponential growth cycle \citep{Pohl2021}.} This was recently shown to be the limiting factor in very young SNRs in dense environments, where the conditions for particle acceleration and turbulence growth were deemed ideal \citep{2021arXiv210813433I}. Additionally, enhanced damping might suppress further turbulence growth when the level of $\delta B/B_0\approx 1$ is surpassed \citep{2021A&A...654A.139B}. In both cases the amplified field can be approximated by a scaling of the existing, ambient field.}

\mponn{Behind the shock, the amplified magnetic field may be quickly damped \citep{2005ApJ...626L.101P}, and so the field would fall back to the large-scale structure that we describe here. }}



\begin{figure}
    \centering
    \includegraphics[width=\linewidth]{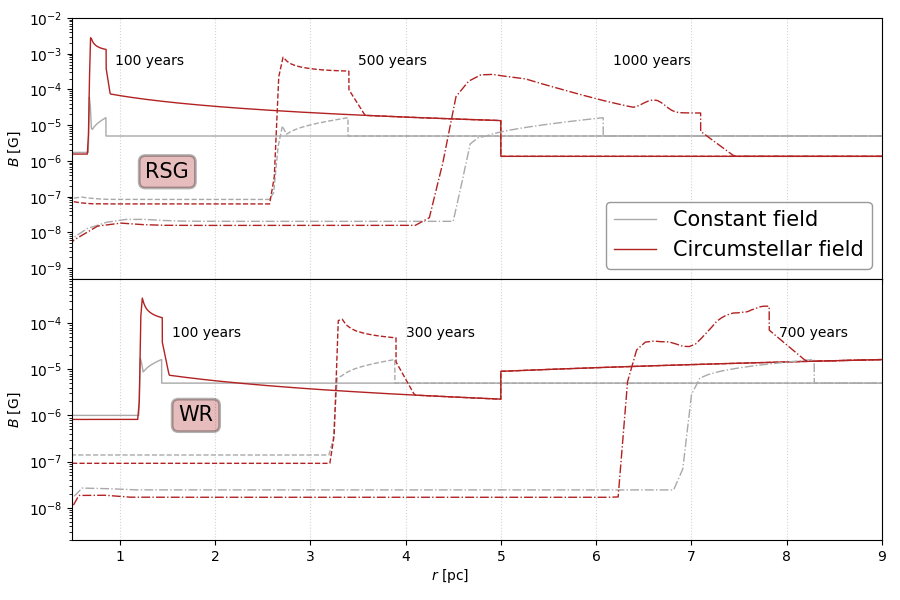}
    \caption{Radial distribution of the magnetic field both downstream and upstream of the forward shock of the SNR at different ages, \mpo{plotted in red. The top panel applies to a RSG progenitor, and the bottom panel is for a WR star.} In both cases an amplified model is assumed and the transition between the \IS{FW} and the \IS{SW} zones is \mpo{located at $r=5$~pc}. These profiles are compared to a simple \mpo{model of a constant $5\,$-$\mu$G upstream} magnetic field, depicted with gray lines in both panels. }
    \label{fig:mf_profiles}
\end{figure}

In Figure~\ref{fig:mf_profiles} we show magnetic-field profiles for the AMP model at different SNR ages for the RSG (top panel) and WR (bottom panel) scenarios. 
Solid lines show the distribution of the magnetic field at a very early stage of the SNR evolution, dashed lines depict the moment shortly before the transition to the \IS{SW} and dashed-dotted lines correspond to the evolution in the SW zone. Gray lines illustrate for reference the profile for a constant ambient magnetic field of 5~$\mu$G. \mpon{The magnetic field is generally stronger for a RSG progenitor than for a WR, and we should expect stronger synchrotron cooling in the SNRs of RSG explosions. Even after the transition to the weaker field in the \IS{SW}, about 1~$\mu$G, the downstream magnetic field remains at the level of several hundreds $\mu$G, as is observed in Cas~A \citep{1980MNRAS.191..855C,2020ApJ...894...51A}. }

As long {\isref as} the SNR evolves inside the FW, we see a similar radial variation in the magnetic-field profiles for the RSG and WR scenarios. The circumstellar magnetic field is 
compressed at the shock and further downstream gradually increases in strength to peak at the contact discontinuity. Upstream of the reverse shock the magnetic field is very weak. 

\mpon{When the SNR shock moves into the \IS{SW}, we see differences between the RSG and WR models, because in the case of a RSG the circumstellar magnetic field becomes weaker, whereas it turns stronger for a WR progenitor.} \is{Moreover, the post-transition phase of the RSG case model is characterized by an additional hump closely behind the shock (about 0.5 pc inward after 1000 years, see top panel of Fig.~\ref{fig:mf_profiles}). This local boost of the magnetic field is caused by compression at a new contact discontinuity (CD) \mpon{between the accelerated forward shock and a reflected shock. This second CD also coincides with a pile-up of particles, and both effects combined lead to additional ring structure in the synchrotron morphology of the source, that we shall} further discuss in Section~\ref{sec:morphology}. }     

%

\subsection{Particle acceleration}

We simulate the evolution of the \mpo{differential particle density, $N$,} by solving the transport equation in the form 
\begin{equation}
\frac{\partial N}{\partial t}=\nabla(D\nabla N-\vec{v}N)-\frac{\partial}{\partial p}\left((N\dot{p})-\frac{\nabla \vec{v}}{3}Np\right)+Q,
\label{Transport}
\end{equation}
where $D$ is the spatial diffusion coefficient, $\vec{v}$ is the plasma velocity, $\dot p$ represents the energy losses (in our case synchrotron losses), and $Q$ is the source term. 

The diffusion is assumed to \mpo{scale with the} Bohm diffusion coefficient
\begin{equation}
  D(p) = \eta_\mathrm{B} \frac{pc^2}{3eB},  
\end{equation}
where $\eta_\mathrm{B}\geq1$ is the Bohm factor and a measure of acceleration efficiency. A recent systematical study of the acceleration efficiency \mpon{suggested a possible} time evolution of the \mpo{diffusion scale factor from $\eta_\mathrm{B}\simeq 10$ at several hundred years to $\eta_\mathrm{B}\approx 1$ after} several thousands years \citep{2021ApJ...907..117T}. The \mpo{implied} extremely high acceleration efficiency obtained for older SNRs is somewhat dubious and fully relies on the assumption that the maximum energy of electrons is limited by synchrotron cooling and the cut-off energy of the X-ray spectrum is independent of the magnetic-field strength \citep{2007A&A...465..695Z}. \citet{2021ApJ...907..117T} motivate this assumption by examining the thin filament-like outer X-ray rims which are supposedly cooling limited.  It was, however, shown already on several occasions that the thickness and brightness of X-ray filaments can be equally well explained by the damping of the magnetic field downstream of the shock, i.e. spatial distribution of the magnetic field, rather than its high strength \citep{2005ApJ...626L.101P, 2012A&A...545A..47R, 2018A&A...618A.155S}. The inaccuracy of the assumption for older SNRs is also supported by estimates of $\eta_\mathrm{B} < 1$ for two SNRs, namely Vela Jr. and HESS J$1731-347$, which \mpo{should be impossible}. 
Therefore, in this study we adopt the value $\eta_\mathrm{B} = 10$ \mpo{for the entire evolution of the SNR up to an age of} 2000 years.  


We use the thermal leakage injection model \citep{2005MNRAS.361..907B}, 
\begin{equation}
Q = \is{\chi} n_\mathrm{u}  (V_\mathrm{sh} - {V_\mathrm{wind}}) \delta(R-R_\mathrm{sh}) \delta(p - p_{\mathrm{inj}}),
\end{equation}
where \is{$\chi$} is the injection efficiency parameter,  
$n_\mathrm{u}$ is the plasma number density in the upstream region, $V_\mathrm{sh}$ is the shock speed, {$V_\mathrm{wind}$ is the wind speed upstream of the shock}, $R_\mathrm{sh}$ is the shock radius, and $p_\mathrm{inj}= \xi p_\mathrm{th}$ is the injection momentum, defined as a multiple of the thermal momentum in the downstream plasma. The injection efficiency for the compression ratio of 4 is determined as
\begin{equation}
  \is{\chi} = \frac{4}{\sqrt{\pi}}\frac{\xi^3}{e^{\xi^2}}.
\end{equation}

We solve the transport equation for electrons in spherical symmetry using the RATPaC code as described in \cite{2012APh....35..300T, 2013A&A...552A.102T, 2018A&A...618A.155S,2019A&A...627A.166B}, taking into account only a forward shock and ignoring a reverse shock. Resulting electron spectra at different moments of time are then used to simulate synchrotron and inverse Compton (scattering on CMB photons) emission from the remnant. In this study we do not consider acceleration of protons and also assume that cosmic-rays will not dynamically affect the SNR evolution.

\section{Results}
\label{sec:results}

\subsection{Evolution in the free wind}

\begin{figure}
    \centering
    \includegraphics[width=\linewidth]{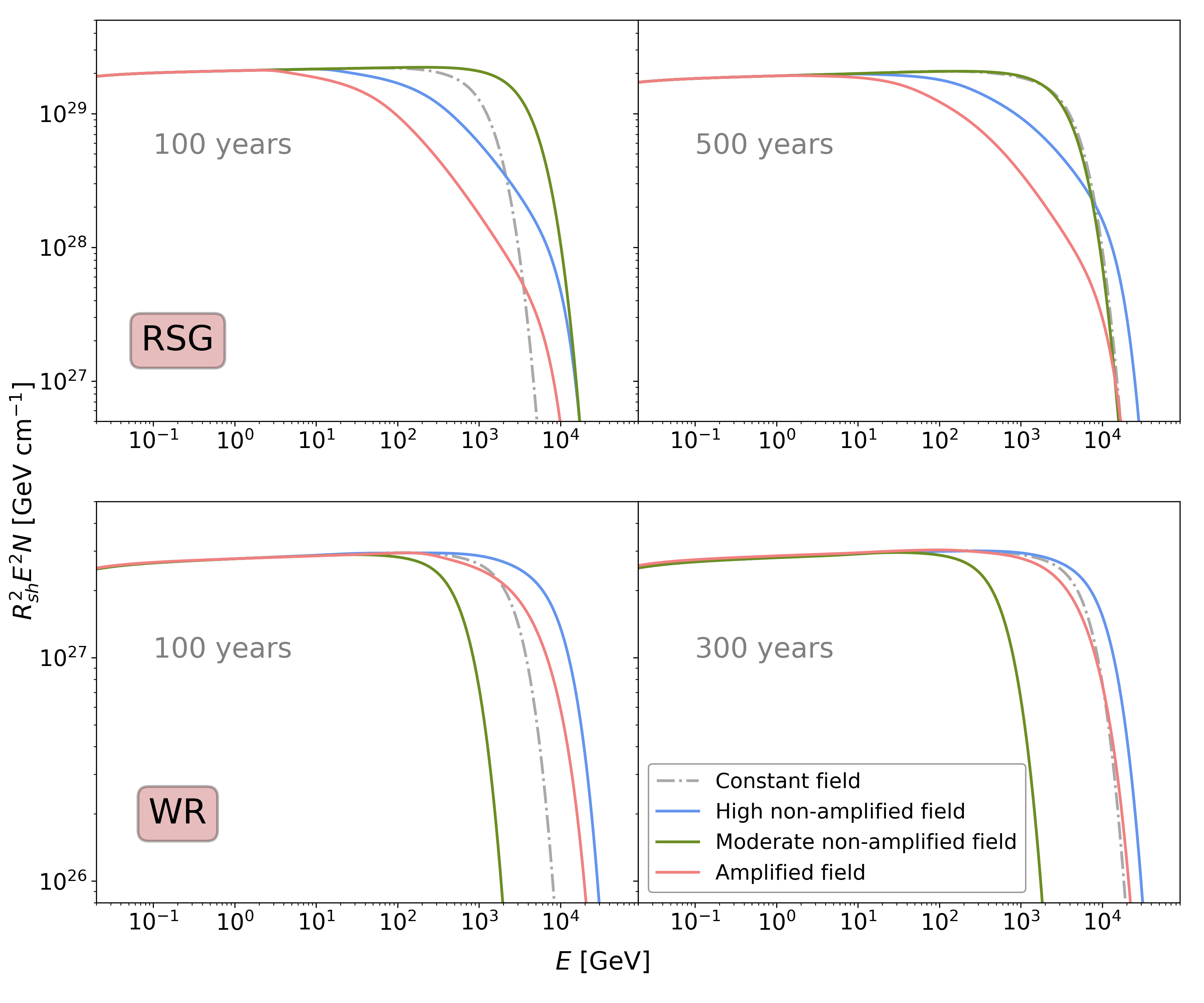}
    \caption{Simulated electron spectra for RSG (top) and WR star (bottom) progenitors at different ages of the SNR. Solid colored lines correspond to different models of the magnetic field in the free-wind zone. The dash-dotted line represents the electron spectrum for a constant ambient magnetic field of 5~$\mu$G.}
    \label{fig:el_spectrum}
\end{figure}

\begin{figure}
    \centering
    \includegraphics[width=\linewidth]{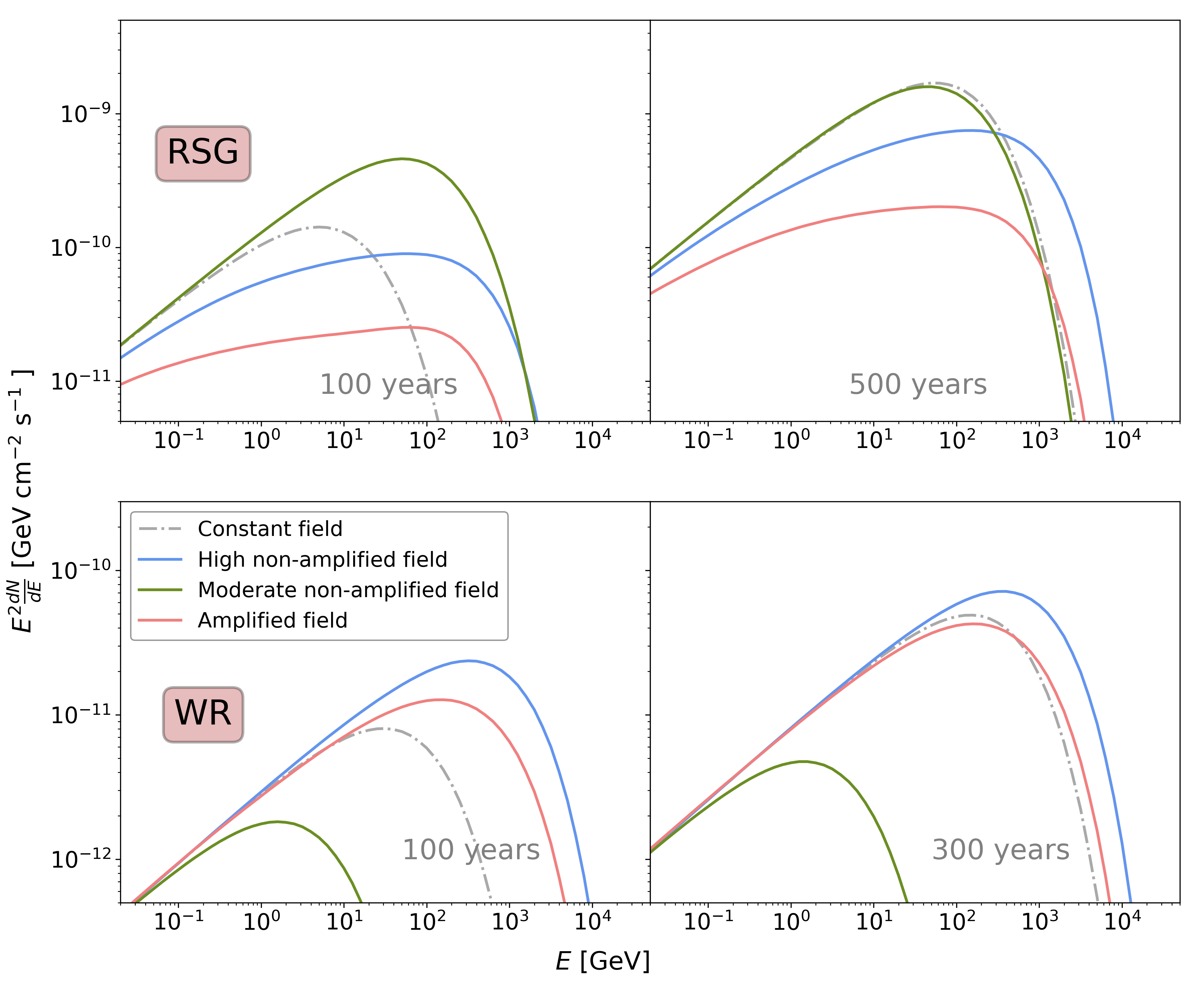}
    \caption{\mpo{Simulated emission spectra for inverse Compton scattering of the CMB, corresponding to the electron spectra shown in Fig.~\ref{fig:el_spectrum}. The style and labeling is preserved.}}
    \label{fig:ic_sed}
\end{figure}

\begin{figure}
    \centering
    \includegraphics[width=\linewidth]{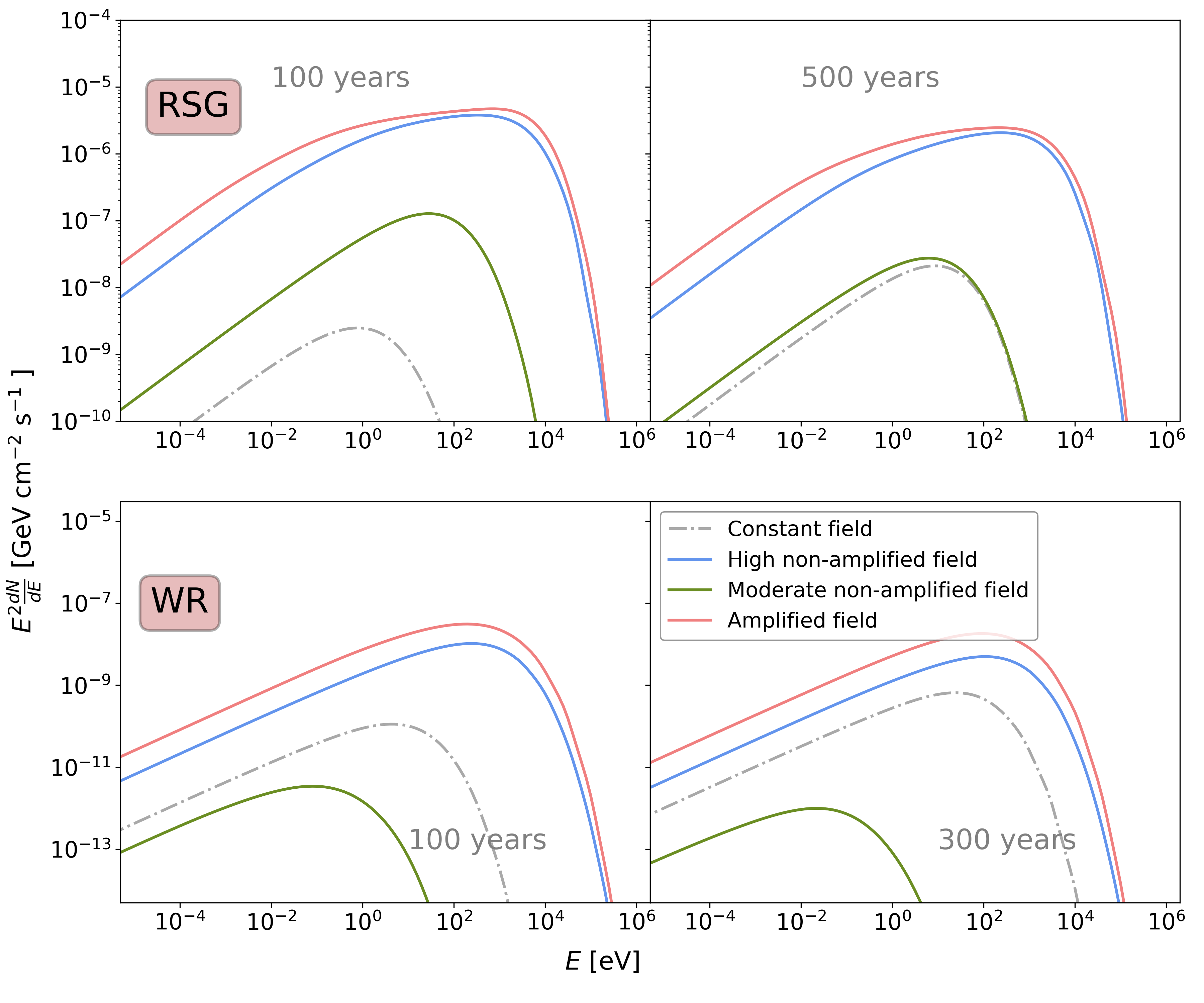}
    \caption{\mpo{Simulated synchrotron spectra corresponding to the electron spectra shown in Fig.~\ref{fig:el_spectrum}. The style and labeling is preserved.}}
    \label{fig:syn_sed}
\end{figure}

First, we simulate the spectra of electrons and the non-thermal emission at the early stage of the SNR evolution, while it still expands in the free wind. Figure \ref{fig:el_spectrum} shows the evolution of the spectrum of accelerated electrons for the four models of the ambient magnetic field for RSG (top panel) and WR (bottom) \mpo{progenitors}. The differential density of electrons is scaled with $R_\mathrm{sh}^2E^2$ for illustration purposes. For reference, the gray dash-dotted line shows the evolution of the electron spectrum for the constant ambient magnetic field of $5$~$\mu$G.

For the RSG progenitor, synchrotron cooling can be substantial even if the magnetic field is not amplified upstream of the shock. The blue line depicts the situation for a strong, but un-amplified circumstellar magnetic field. The electron spectrum features a cooling break at a few tens of GeV after 100 years, that eventually moves to higher energies as the field strength decreases. 
Already for the moderate magnetic field the cooling is inefficient and does not significantly modify the spectrum (green line), \mpo{but the $r^{-1}$ distribution of the magnetic field in the free-wind zone can still leave imprints in the} time evolution of the maximum electron energy. \mpo{For a} constant ambient magnetic field, the maximum energy is age limited, hence it evolves as $E_{\rm max}^{\mathrm{age}} \propto V_{\rm sh}^2  t B$, where $V_{\rm sh}$ is the shock velocity \citep[see e.g.][]{2008ARA&A..46...89R}. 
\is{At the free expansion stage the shock radius \mpo{evolves} as $R_\mathrm{sh}\propto t^{(n-3)/(n-m)}=t^{6/7}$ \citep{1999ApJS..120..299T}, where the index of the ejecta density profile is $n=9$ and the index of the density profile in the wind zone is $m=2$. The shock speed \mpo{declines} as $V_\mathrm{sh}\propto t^{-1/7}$, yielding $E_{\rm max}^{\mathrm{age}}\propto t^{5/7}$ for the constant magnetic field, whereas for \mpo{the realistic scaling,} $B\propto r^{-1}$, we expect $E_{\rm max}^{\mathrm{age}}\propto t^{-1/7}$.} 

For \mpo{the strong circumstellar magnetic field encountered} during the early stages of SNR evolution, the maximum energy is limited by synchrotron cooling. \mpo{The balance of cooling and acceleration yields $E_{\rm max}^{\mathrm{syn}} \propto V_{\rm sh} B^{-1/2}$ \citep[see e.g.][]{2008ARA&A..46...89R}, and hence the maximum energy} 
\is{only slowly grows with time, as $E_{\rm max}^{\mathrm{syn}} \propto t^{2/7}$.}

For the WR progenitor star (Fig.~\ref{fig:el_spectrum}, bottom panel), synchrotron cooling is much less efficient even for \mpo{a very strong} magnetic field. Non-amplified field models result in featureless spectra roughly \mpo{similar to those for a constant, weak} ambient field. The amplified field model still shows \mpo{a cooling break at a few hundred GeV early in the evolution, but already 300 years after the Supernova the electrons \is{\mpo{have a} featureless uncooled spectrum.}} \mpo{The cut-off energy for the circumstellar magnetic field does not significantly increase with time, unlike that for the constant ambient field, indicating that it is still limited by synchrotron cooling.}

Figure \ref{fig:ic_sed} shows the inverse Compton emission generated \mpo{by CMB radiation scattering off the electrons. The gamma-ray spectra of SNRs resulting from the explosion of RSGs can feature a low-energy break and display spectral softening above the break. The spectral break appears below the GeV scale} at early stages and gradually shifts towards higher energies, as the ambient magnetic field decreases. The spectral shape visually resembles the one expected for \mpo{hadronic gamma-ray emission, introducing new difficulties in discrimination between the leptonic and hadronic scenarios, but it is much smoother then the hard transition in the pion-decay spectra, emphasizing the need for reliable data below $500~$GeV}.

For the WR progenitor (Fig.~\ref{fig:ic_sed}, bottom panel), \mpo{we find only a weak spectral modification, and that only for the amplified field. By and large,} there is no significant change of the spectral shape as compared to the constant low ambient magnetic field.

\is{Similarly, the simulated synchrotron spectra (Fig.~\ref{fig:syn_sed}) reveal a characteristic cooling break for the \mpo{strong-}field configurations in the RSG scenario, \mpo{but nothing of that kind} in the WR scenario. The \mpo{flux normalization} for circumstellar field models decreases with time, \mpo{unlike for the} constant ambient field, which can be directly attributed to the decrease of the magnetic field itself. This would naturally result in the decrease of the expected radio and X-ray emission, which is further discussed in Section~\ref{sec:lc}.}

\subsection{Transition to the shocked wind medium}

\begin{figure}
    \centering
    \includegraphics[width=\linewidth]{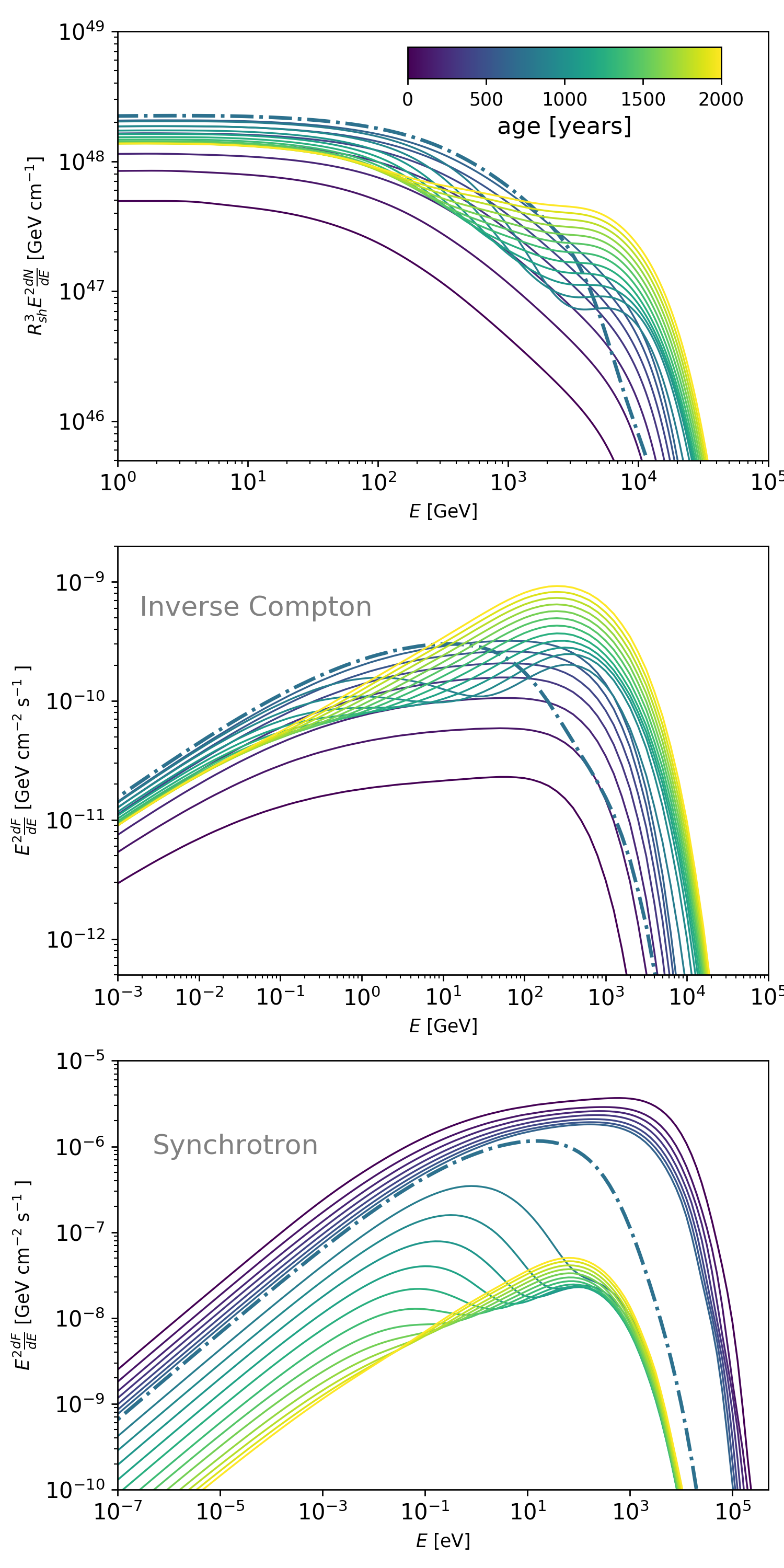}
    \caption{Volume-averaged electron spectrum normalized with $R_\mathrm{sh}^3$ (top panel) and the corresponding gamma-ray \mpo{(middle panel) and synchrotron emission spectra (bottom panel) for a RSG progenitor. \mpo{The spectra are color-coded for SNR age with increments of $100$ years. We highlight with a dash-dotted line the spectrum close to the transition from the free RSG wind to the main-sequence wind, when the age is around} $800$ years}.}
    \label{fig:RSG_jump}
\end{figure}    

\begin{figure}
    \centering
    \includegraphics[width=\linewidth]{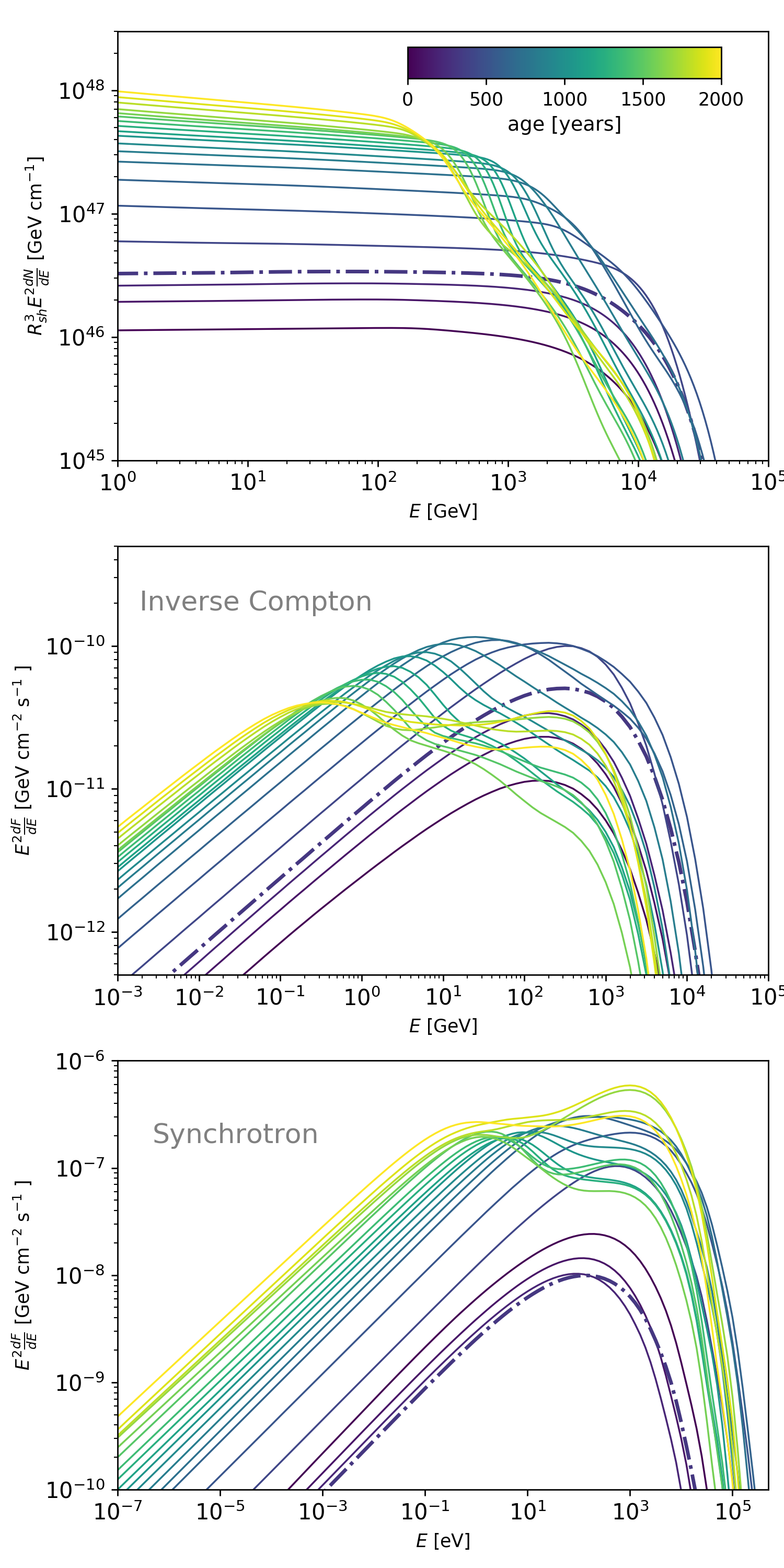}
    \caption{Same as Fig.~\ref{fig:RSG_jump} but for the WR progenitor. \mpo{The dash-dotted line gives the spectra close to the transition to the \IS{SW} after around $400$ years}. 
    }
    \label{fig:WR_jump}
\end{figure}

The transition to the \IS{SW} zone is accompanied by an abrupt change of the physical properties of the medium. To investigate the impact of the change of magnetic field on the resulting particle and gamma-ray spectrum, we 
use AMP models described in Section~\ref{sec:CM}. 

The evolution of the electron, gamma-ray and synchrotron spectra \mpo{for a} RSG progenitor is shown in Fig.~\ref{fig:RSG_jump}. 
\is{Before the transition to the \IS{SW} \mpo{after nearly $800$~years}, the evolution of the electron spectrum is governed by synchrotron cooling and the linear decrease of the magnetic field strength. As the remnant expands, the cooling break shifts to higher energies, while the maximum energy of electrons slowly increases (see previous Section). \mpo{After $800$ years, the dash-dotted spectrum shows a significant shift of the cut-off to lower energies which reflects the entrance of the cosmic-ray precursor into the weak-field region beyond the transition point. The diffusion coefficient in the precursor increases, and high-energy particles can effectively} escape. As soon as the shock enters the dilute shocked medium, the shock velocity increases considerably, and particles are accelerated to high energies again, despite the weak magnetic field. }
Already after $900$ years a second very-high-energy component emerges in the electron spectrum, which becomes more significant with time and eventually starts to dominate (Fig.~\ref{fig:RSG_jump} top panel). \is{In the IC spectrum, a hint of this component could be seen already after $800$ years (middle panel).} 
\mpo{Whereas some of the high-energy particles close to the shock can receive a further boost in energy, low-energy particles get advected downstream, escape the acceleration process, and are subjected to synchrotron cooling}. This explains the softening of the spectrum above the cooling break energy at a few hundred GeV \is{shortly after the transition}. \is{Later on, the spectrum in this energy range gets harder again due to the rise of the second component.} Reaccelerated particles together with freshly accelerated particles constitute \is{this} very-high-energy component of the spectrum. Moreover, \mpo{cooling becomes ineffective close to the shock} due to the much lower magnetic field in the shocked medium. The resulting concave spectrum can be described as a superposition of spectra of two electron populations: old cooled electrons with a spectral break at around 100 GeV and new uncooled electrons with no cooling break and higher maximum energy. With time, \mpo{further cooling of the first population shifts the break energy to lower energies, and the younger particles become more dominant due to their increasing number}. This gradually changes the concavity of the spectrum until at some point in time the spectrum becomes featureless again.

The concavity of the electron spectrum is reflected in the radiation spectra (Fig.~\ref{fig:RSG_jump},  middle panel for inverse Compton scattering and bottom panel for synchrotron), strongly changing their shape in the \mpo{GeV band and the keV band}. The \mpo{appearance of the new} high-energy component in the electron spectrum causes a strong temporal variation of the spectral shape in the GeV-to-TeV band. Right after the transition to the \is{shocked} \mpo{main-sequence wind} the very-high-energy spectrum is soft, \mpo{but gradually hardens, ending up with a typical featureless IC spectrum when the new} high-energy component takes over completely. \mpo{This evolution takes} around $700$ years for our setup, suggesting that \mpo{the different phases of spectral evolution} could potentially be detected. Although the main reason for the concavity of the spectrum is the abrupt change of the magnetic field, the change in density would regulate the timescale on which the gamma-ray spectrum \mpo{settles to the typical form}. The higher the amplitude of the density drop between the \mpo{RSG and the main-sequence wind, the slower is the evolution of the spectrum}, due to the lower injection rate of particles at the shock.

Similar variations are seen in the synchrotron spectrum, the main difference lying in the duration of the process. Since the synchrotron radiation directly depends on the magnetic-field \mpo{strength}, the new high-energy component takes longer \mpo{to emerge in the synchrotron spectrum} than in the inverse Compton spectrum. Additionally, the normalization of the spectrum decreases \mpo{as the magnetic-field strength drops}.


The situation with the WR progenitor is quite the opposite \is{and more complicated}, \mpo{but eventually leads} to a similar outcome\is{, i.e. a concave spectrum} (Fig.~\ref{fig:WR_jump}). \is{The magnetic field generated by the WR progenitor is weaker than that generated by the RSG progenitor, implying that synchrotron cooling is less efficient. Indeed, only at very early times a hint of a cooling break can be seen, and it disappears \mpo{quickly}. Nevertheless, the maximum energy of electrons is still \mpo{cooling-limited}, implying a slow rise of the maximum energy (see the previous section). }

\is{As the forward shock approaches the transition point after nearly $400$ years, \mpo{as in the RSG case the cosmic-ray precursor extends to the \IS{SW} and the stronger magnetic field there, leading to a better confinement. 
The forward shock instantaneously decelerates by a factor of $1.5$, when it enters the denser \IS{SW}, allowing particles of the highest energy to return} to the shock and re-enter the acceleration process. \mpo{A prominent effect is the sharpened cut-off that is} seen after $500$ years. As the remnant propagates through the \IS{SW} with the linearly increasing magnetic field, synchrotron cooling becomes more efficient \mpo{and imposes a cooling break and a spectral} softening at high energies. Particles \mpo{located sufficiently far downstream still see the lower magnetic field and hence weak cooling} (see Fig.~\ref{fig:mf_profiles}), but they also have a lower maximum energy. }
This again creates a two-component particle spectrum \mpo{and} concave inverse Compton and synchrotron spectra. \is{With time the strong magnetic field encountered in the \IS{SW} advects to \mpo{the far downstream and renders synchrotron cooling relevant also for} the low energy component.} \mpo{Like for the RSG progenitor}, \is{eventually} the high-energy component becomes dominant, allowing for \mpo{considerable evolution} of the spectral hardness in the GeV-TeV and eV-keV bands. This \mpo{spectral modification is slower than for a RSG progenitor, opening an even larger window of opportunity for observing it}. As expected, the maximum energy of accelerated electrons slowly decreases with time, as the shock slows down and the magnetic field \is{increases linearly}.

The 
interaction with the termination shock can \mpo{briefly increase the compression ratio felt by particles and hence induce a spectral bump \citep{2013A&A...552A.102T}, but the duration of this event is too small to give} a significant impact \is{\citep{Das:20216k}}, and therefore it is ignored here. More importantly, such a \mpo{shock collision would generate a reflected shock propagating in a medium of lower density that might eventually be reflected off the contact discontinuity of the SNR and collide with} the forward shock. This process might repeat several times. The corresponding boost of the shock speed results in an instant increase of the maximum electron energy, modifying the spectrum accordingly. \is{This would result e.g. in brightening and hardening of the X-ray and TeV emission on very short time scales. The recent detection of rapid variations in X-ray filaments of the Tycho's SNR \citep{2020ApJ...894...50O} may be attributed to the boosting of the forward shock \mpo{by a trailing shock. A second possible explanation is dynamic magnetic turbulence \citep{2008ApJ...689L.133B}.}}
After the \mpo{boosting event} the shock decelerates again, and the maximum energy decreases until the next interaction happens. \is{The reflected shocks bouncing back and forth between the forward shock and the CD would compress and re-accelerate particles, which can further modify the spectrum and the source morphology. \mpo{For the WR progenitor} the reflected shock is responsible for somewhat irregular variations of the electron and radiation spectra \mpo{that are superposed on the} main trend determined by the change of the magnetic field.} 




\subsection{Time evolution of the non-thermal emission}
\label{sec:lc}

\begin{figure}
    \centering
    \includegraphics[width=\linewidth]{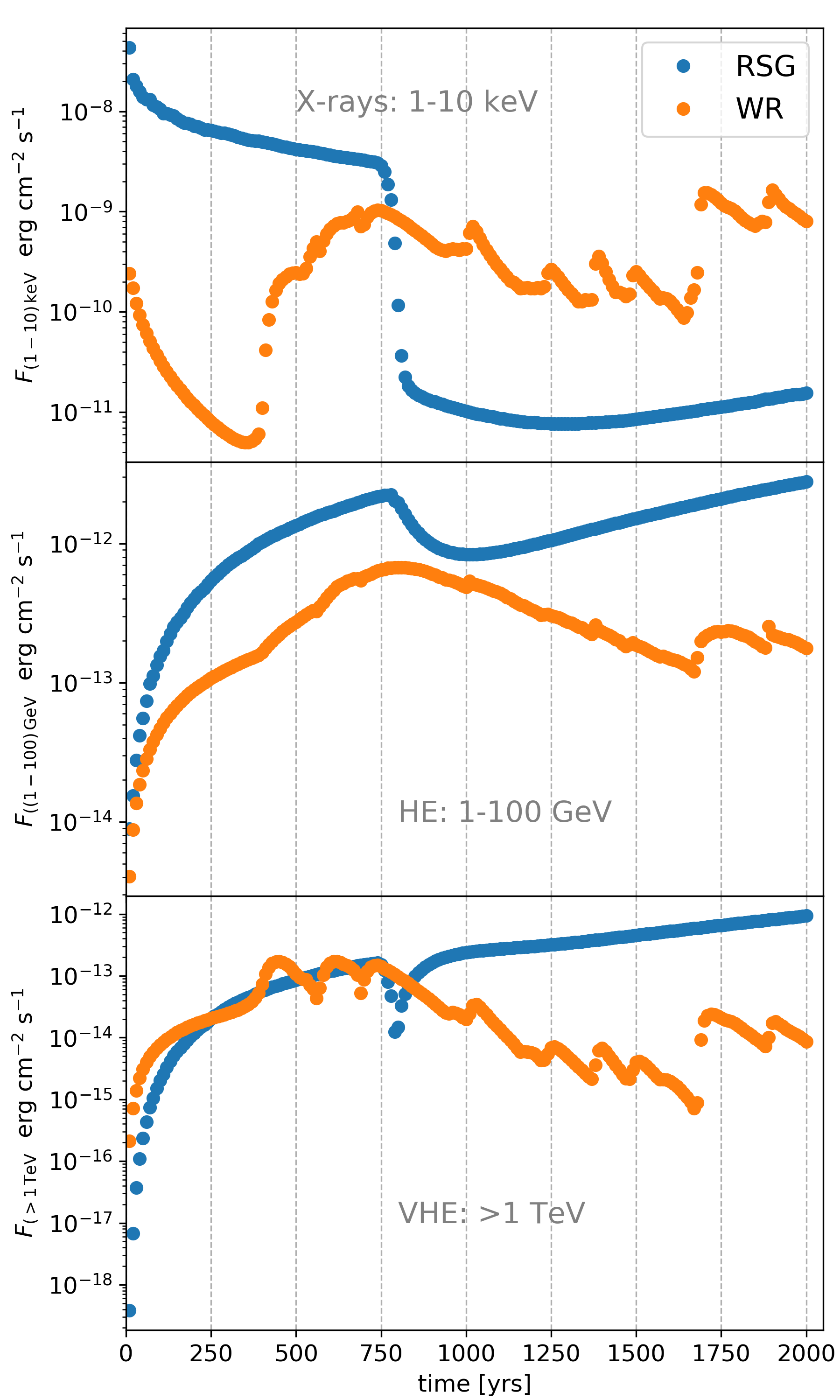}
    \caption{\is{Simulated light curves for X-rays \mpo{at $1-10$~keV (top), the $1-100$~GeV band (middle), and above a TeV. Blue filled circles apply to the RSG progenitor, while the orange circles are for} the WR scenario.}}
    \label{fig:lc}
\end{figure}

\mpo{In Fig.~\ref{fig:lc} we show the simulated flux in X-rays, high-energy gamma-rays, and very-high-energy gamma-rays for our AMP models and RSG and WR progenitors}. At the early stages of evolution (in the FW zone) both \mpo{scenarios, RSG and WR, feature a power-law decrease of the X-ray flux, albeit with} different slope. For the RSG progenitor, the dimming of the X-ray emission is determined exclusively by the $1/r$ scaling of the magnetic field strength. The \mpo{decreasing upstream density, and likewise the injection rate, is fully compensated by the increasing shock surface, because the synchrotron cut-off does not fall into the energy range in question.} 
For the WR progenitor the \is{expected X-ray flux is sensitive to the change of the cut-off energy of the radiation spectrum, which depends on the maximum electron energy and magnetic field as $E_\mathrm{cut}^\mathrm{ph} \propto E_\mathrm{max}^2B$ \citep{2008ARA&A..46...89R, 2012A&ARv..20...49V}. For the synchrotron-limited particle acceleration in the FW zone the cut-off energy \mpo{evolves} as $E_\mathrm{cut}^\mathrm{ph}\propto t^{-2/7}$.}  
\mpo{A consequence is} faster dimming of the X-ray emission than in the RSG case. \mpo{When the shock leaves the FW zone,} the X-ray flux abruptly decreases in the case of the RSG and increases for the WR \mpo{progenitor, on account} of the \mpo{jump of the magnetic field which besides the synchrotron emissivity also changes} \is{the confinement of high energy particles at the shock}. In the WR scenario the flux keeps increasing for a few hundred years, which is partly due to re-acceleration of old particles to higher energy and partly due to two episodes of \mpo{shock-shock} interaction. These episodes happen at around $560$ years and $690$ years and are \mpo{clearly reflect}ed in the VHE light curve (Fig.~\ref{fig:lc} bottom panel). Later, \mpo{the forward shock decelerates, causing a gradual decline of the X-ray flux} \is{enriched with sporadic brightenings caused by numerous shock-shock interactions}.

Unlike synchrotron radiation, the gamma-ray flux generated by inverse Compton scattering \mpo{simply reflects the number of electrons and does not directly depend on the magnetic field strength. As long as the SNR expands} into the FW, the HE and the VHE flux grow, because more and more particles get injected into the acceleration process. \mpo{Like the X-ray flux, the gamma-ray flux abruptly drops at the transition to the main-sequence wind in the RSG scenario. This decline is mainly caused by synchrotron cooling of electrons accelerated in the FW zone that cannot be replaced on account of the abruptly decreased injection. \mpo{In line with the energy dependence of the cooling time, the TeV-band flux recovers much faster than the GeV-scale flux}. Later on the flux increases in both energy bands, reflecting the continuously growing number of accelerated particles.} 

For the WR \mpo{progenitor we observe} the opposite. At the transition to the \IS{SW}, the \mpo{flux in the entire gamma-ray band is boosted by the higher injection of particles and keeps growing until shock deceleration and synchrotron cooling introduce a slow decline}, which, naturally, happens earlier in the VHE range than in the HE range. The VHE light curve also features several \mpo{bumps, which reflect episodes of shock-shock interaction. The boost of the forward shock} instantaneously increases the maximum energy of accelerated particles and hence the VHE flux. \mpo{The counterparts of these bumps in the GeV light curve are much less prominent, if at all visible.}  


\is{A \mpo{declining flux of non-thermal X-rays was observed from Cas~A \citep{2011ApJ...729L..28P} a remnant of the Type-IIb explosion of a} RSG \citep{2008Sci...320.1195K}. The $4.2-6$~keV flux from the whole remnant decreased at a rate of $1.5\%/\mathrm{yr}$ over 11 years which the authors attributed to a declining maximum energy due to shock deceleration. They argued that the decrease of the magnetic field alone would not be sufficient to explain such a decline \citep{2011ApJ...729L..28P}. The required deceleration rate of the forward shock considerably exceeds the observed deceleration with the expansion parameter of $q\simeq0.65$ for $R_\mathrm{sh}\propto t^{q}$ \citep{2009ApJ...697..535P}. Note that this expansion parameter is compatible both with the Sedov-Taylor stage of an SNR in the FW, $q = 2/(5-m) = 2/3$ with $m = 2$ for a $\rho\propto r^{-m}$ density profile, and with the free expansion of a core-collapse SNR in a medium of constant density, $q=(n-3)/(n-m) = 2/3$ for $n=9$ and $m=0$. The latter reflects exactly the situation that we expect for the RSG scenario after the transition to the shocked-main-sequence wind. The expansion of the remnant is still ejecta-dominated as it expands in a very dilute medium, and the density of the \IS{SW} is roughly constant. Interestingly, the average dimming rate of the X-ray emission near the transition time in our simulations, between $770$ and $840$ years, is $1.4\%/\mathrm{yr}$, compatible with \mpo{that} observed for Cas A. In our simulation the dimming arises from the escape of the high-energy particles upstream of the shock, where the magnetic field abruptly decreases. For about $10$ years during the transition phase the dimming can reach $ 10\%/\mathrm{yr}$, but as soon as the \mpo{forward shock enters the \IS{SW}, the dimming normalizes to $1-1.5\%/\mathrm{yr}$. At this time, the shock expands in a constant-density environment and thus decelerates again. Note that deviations from spherical symmetry can smear out this effect over time.} It is plausible that Cas~A at its age around $350$ years recently transitioned from the FW to the \IS{SW} of its progenitor stellar wind bubble, and the decline of the X-ray emission that we observe \mpo{reflects} this transition.} 

\is{\mpo{The rate of change in the gamma-ray flux is smaller} but can still reach a few $\%/\mathrm{yr}$ near the transition time. Such a flux change \mpo{is potentially detectable with current instrumentation} if we observe the source at the right moment. \mpo{For WR progenitors, sporadic brightenings arising from shock-shock} interactions are also strong enough to be detected. \mpo{An identification as shock interaction events is likely impossible though, because their duration is longer than the time period available for observations. The general brightening in the TeV band in the RSG case, \IS{$ \sim 0.25\%/\mathrm{yr}$} beyond the transition point, and likewise the dimming for the WR scenario, \IS{$\sim-0.15\%/\mathrm{yr}$,}} are too slow to be currently detectable.}



\subsection{Morphology}
\label{sec:morphology}

The two \mpo{spectral components of electrons, that we saw when the \mpo{SNR shock left the FW,} are not only spectrally distinct} but also spatially separated. The pre-transition population of electrons, that were accelerated \mpo{when the shock propagated through the FW zone}, are located further inside the remnant, whereas the freshly accelerated, more energetic electrons are found close to the shock. \mpo{In addition, the speed of the forward shock is modified, changing the separation between the forward shock and the CD, which would likewise} be reflected in the morphology. This should \mpo{be particularly prominent} in the radio emission, as low-energy \mpo{electrons preferentially reside close to} the CD, and the magnetic field piles up there \citep{1976PhFl...19.1889R}.

\begin{figure}
    \centering 
    \includegraphics[width=\linewidth]{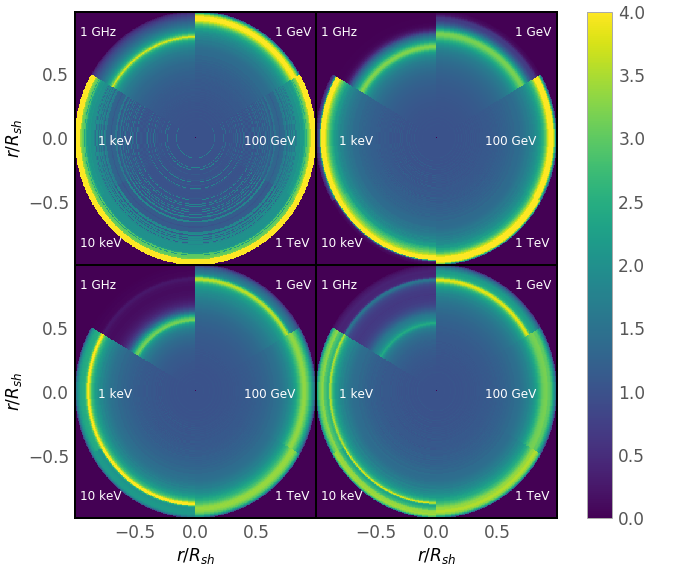}
    \caption{\mpo{Simulated intensity maps for a RSG progenitor at the ages $500$~years (top left), $900$ years (top right), $1500$ years (bottom left), and $2000$ years (bottom right). Each sectors displays the projected brightness at a specific energy, normalized to that at the center. The left hemisphere exhibits synchrotron radiation at $1$~GHz (top), $1$~keV (middle), and $10$~keV (bottom), and the right hemisphere shows IC emission at $1$~GeV (top), $100$~GeV (middle), and $1$~TeV (bottom).}}
    \label{fig:mapsRSG}
\end{figure}

\begin{figure}
    \centering
    \includegraphics[width=\linewidth]{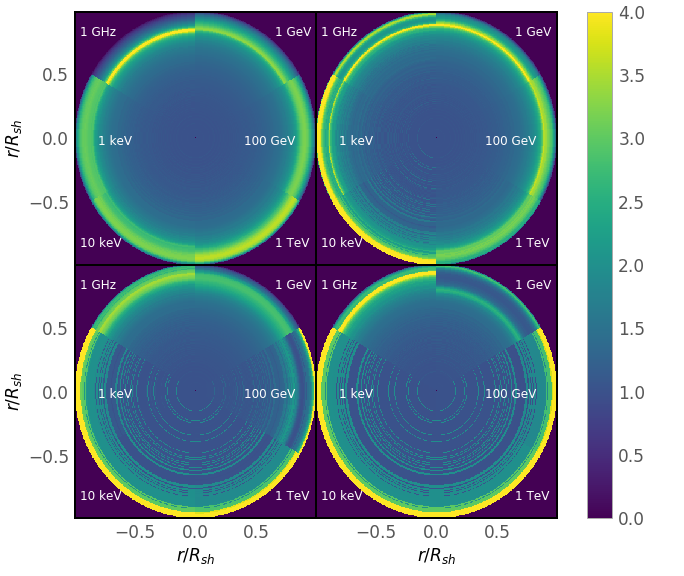}
    \caption{Same as Fig.\ref{fig:mapsRSG}, but for the WR progenitor and the ages $300$~years (top left), $500$ years (top right), $1000$ years (bottom left), and $2000$ years (bottom right).} 
    \label{fig:mapsWR}
\end{figure}

In Figs.~\ref{fig:mapsRSG} (RSG) and~\ref{fig:mapsWR} (WR) we show simulated intensity maps for different ages and different photon energies. The top left panels depict the morphology \mpo{when the SNR shock propagates through the FW zone. Both X-ray and TeV-scale emission are bound to the location of the shock where the high-energy particles reside. To be noted from the figures is that t}he radio shell is located farther \mpo{at} the CD where the low-energy particles pile up and the magnetic field is high. \mpo{Time-variable X-ray filaments have been observed in the interior of Cas~A \citep{2008ApJ...677L.105U} and in a number of other SNRs. They may be caused by the strong field at the CD, if the site of electron acceleration is very close \citep{2004ApJ...609..785L}. Our simulation assumes that most of the acceleration happens at the forward shock, and so we do not see X-ray features associated with the CD. }

\is{In the WR scenario \mpo{the rim of $1$-GeV emission is roughly co-located with the radio shell, reflecting the high density of low-energy electrons there.} In the RSG case, however, the GeV shell \mpo{is closer to the forward shock, on account of the strong synchrotron cooling of the TeV-scale electrons that produce the GeV-scale IC emission, whereas} the radio-emitting GeV-band electrons remain unaffected.} 

\is{In the RSG case, once the remnant transitions to the shocked medium, the forward shock speeds up. The separation between the shock and the CD increases, \mpo{and so does that between the inner radio ring and the outer X-ray shell.} The acceleration of the forward shock also results in the offset of the $1$~GeV ring towards the interior of the remnant, \mpo{because the production rate of TeV electrons has collapsed.} Right after the transition, after $900$ years (upper right map of Fig.~\ref{fig:mapsRSG}), the \mpo{highest X-ray and very-high-energy gamma-ray intensity is still seen close} to the shock. At later times a second CD is formed, where particles are piled up and the magnetic field is compressed (Section~\ref{sec:Bfield_model}). \mpo{After $1500$ years,} synchrotron cooling is much less effective, and the GeV-TeV morphology is similar to \mpo{that seen during the early phase in the WR case}, i.e. a shell location gradually shifting from the CD for lower energies to the forward shock for higher energies (bottom left map on Fig.~\ref{fig:mapsRSG}). The formation of the second CD also results in the \mpo{appearance of a} second radio ring that is prominent after $2000$ years (bottom right map on Fig.~\ref{fig:mapsRSG}). Initially, \mpo{high X-ray intensity coincides with} this second CD on account of the higher magnetic field, but later, after $2000$ years, a second X-ray shell emerges that \mpo{is caused by} freshly accelerated particles at the shock.}

\is{For the WR progenitor, the transition to the shocked medium immediately results in a double-ring structure in the radio and soft X-ray morphology (top left map of Fig.~\ref{fig:mapsWR}). These two rings reflect the locations with enhanced magnetic field: the CD and the abruptly increased field at the shock. Later on, as the magnetic field and particles are advected downstream, the synchrotron morphology is mainly determined by the spatial distribution of electrons downstream. Note, that although the radio emission is generated farther in the interior of the remnant, it no longer \mpo{coincides with the CD, because we do not find the highest electron density and strongest magnetic field there}. At later stages of evolution (bottom maps on Fig.~\ref{fig:mapsWR}), a second ring of \mpo{GeV-band emission is formed by the electrons near the CD that gradually cool} with time. After $1000$ years (bottom left map on Fig.~\ref{fig:mapsWR}) the inner ring appears in the $100$-GeV map, and after $2000$ it is most prominent at $1$~GeV.}

\begin{figure}
    \centering 
    \includegraphics[width=\linewidth]{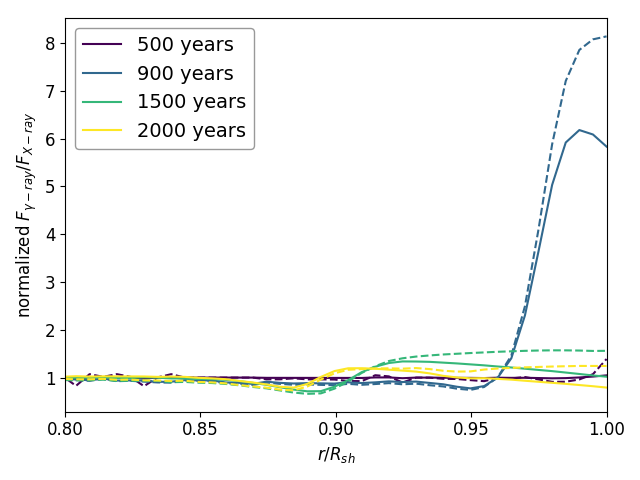}
    \caption{\mpon{Intensity ratio of gamma rays and X rays for the RSG scenario, plotted as a function of the projected distance from the center and normalized to unity at the center. The X-ray brightness is integrated between 1 and 5 keV, whereas for the gamma rays we use two energy bands: above 250 GeV (solid lines) and above 1 TeV (dashed lines). The line color indicates the age of the remnant.}}
    \label{fig:fluxratioRSG}
\end{figure}

\begin{figure}
    \centering
    \includegraphics[width=\linewidth]{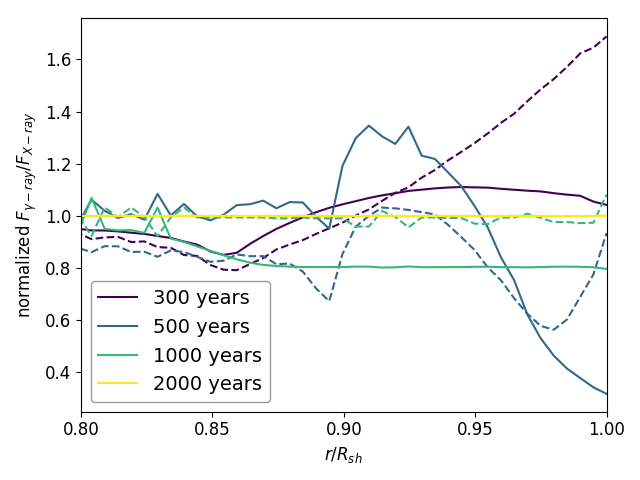}
    \caption{\IS{Same as Fig.\ref{fig:fluxratioRSG}, but for the WR progenitor.}}     \label{fig:fluxratioWR}
\end{figure}

\IS{X-ray and IC very-high-energy gamma-ray emission are produced by electrons of roughly the same energy. It is therefore tempting to assume that both emissions should be produced in the same regions \mpon{and with proportional emissivity}, hence the gamma-ray morphology would strongly correlate with the X-ray morphology. \citet{2021ApJ...915...84F} \mpon{made this assumption} to discriminate between the leptonic and hadronic scenario of the gamma-ray emission, \mpon{likewise assuming} that hadronic gamma-ray radiation should correlate with the gas distribution. \mpon{The latter assumption is dubious because we measure column densities and not the gas density itself, that determines the hadronic emissivity. Besides, the intensity maps in Figs.~\ref{fig:mapsRSG} and \ref{fig:mapsWR}} show clearly that the X-ray and the leptonic gamma-ray morphology do not necessarily correlate, as the former is also strongly dependent on the \mpon{local} magnetic field. To illustrate this in more detail we calculate projected radial profiles of the gamma-ray to X-ray flux ratio (Figs.~\ref{fig:fluxratioRSG} and \ref{fig:fluxratioWR}). Following \citet{2021ApJ...915...84F}, we calculate the gamma-ray flux in two energy bands: above 250 GeV (solid lines) and above 1 TeV (dashed lines). The X-ray flux is calculated in the energy range of $1-5$~keV. \mpon{For both types of SNR we do not see a constant flux ratio. Instead}, a clear anti-correlation of the gamma-ray and X-ray morphology is visible for certain ages of the SNR at length scales of $5-15\%$ of the shock radius. For larger SNRs, this could be well resolved even by current generation instruments.}

\is{
\subsection{Spatial variation of the GeV spectrum}}
\begin{figure}
    \centering
    \includegraphics[width=\linewidth]{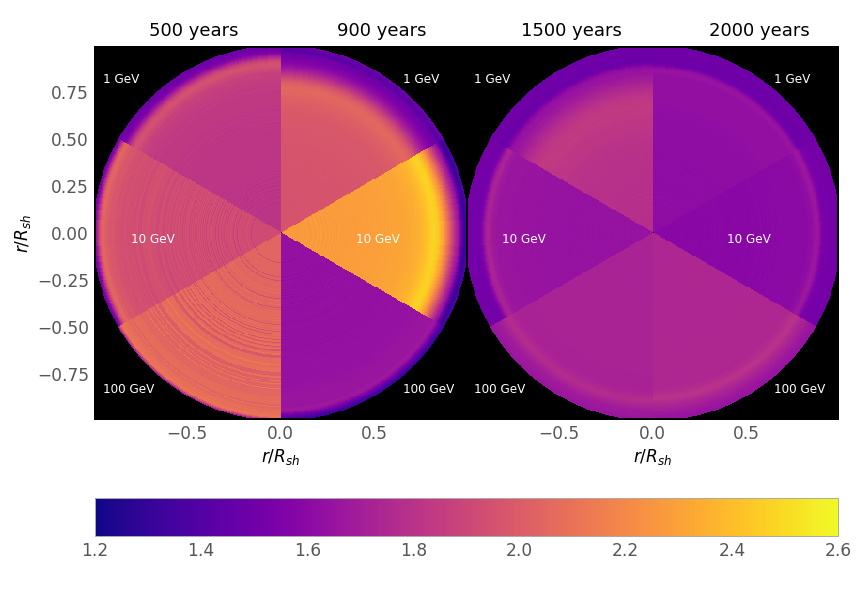}
    \caption{\is{Photon spectral index maps for 1~GeV (top sector), 10 GeV (middle sector), and 100 GeV (bottom sector) calculated for the RSG \mpo{scenario at the same ages as in Fig.~\ref{fig:mapsRSG}}.}}
    \label{fig:indRSG}
\end{figure}

\begin{figure}
    \centering
    \includegraphics[width=\linewidth]{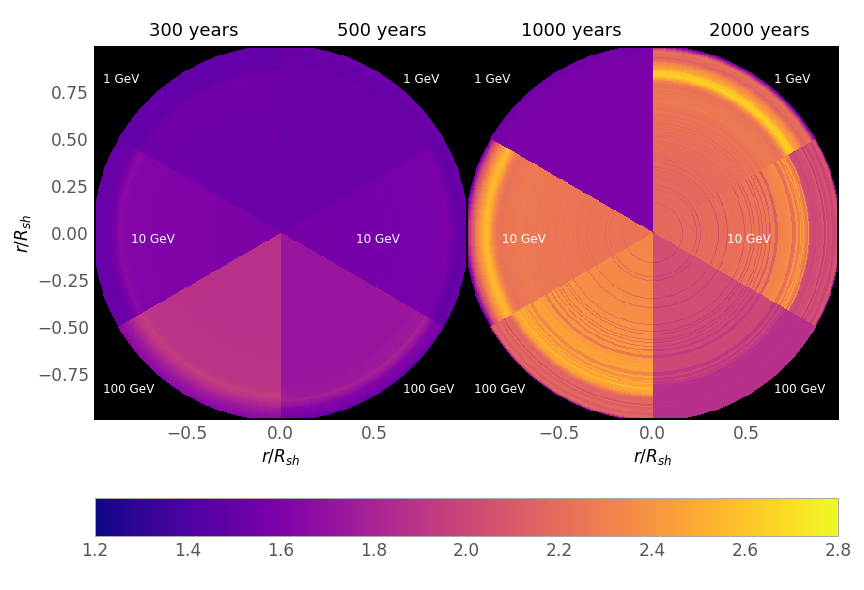}
    \caption{\is{Same as Fig.~\ref{fig:indRSG}, but for the WR case and the ages as in Fig.~\ref{fig:mapsWR}.}}
    \label{fig:indWR}
\end{figure}

\is{As already emphasized in the previous subsection, two \mpo{spatially separated populations of electrons are seen after} the transition of the forward shock to the shocked medium. \mpo{We calculated spectral index maps at three different energies, $1$~GeV, $10$~GeV, and $100$~GeV, to search for corresponding spatial variations of the spectral index, which may be relevant} for the discrimination between leptonic and hadronic scenarios. \mpo{We chose the same ages as for} the emission maps in Fig.~\ref{fig:indRSG} for the RSG case and Fig.~\ref{fig:indWR} for the WR progenitor. The index plotted is the photon spectral index, $\alpha$, defined as}
\is{
\begin{equation}
    N_\gamma(E)=\frac{dN}{dE}\propto E^{-\alpha},
\end{equation}
where $N_\gamma(E)$ is the differential photon flux. At each location the index for particular energy $E$ is calculated as
\begin{equation}
    \alpha(E)= \log \frac{N_\gamma(E-dE)}{N_\gamma(E+dE)} \Bigg/ \log \frac{E+dE}{E-dE}
\end{equation}}

\is{During the pre-transition stage of the RSG \mpo{scenario, the GeV-band emission spectrum is flat in $E^2N_\gamma(E)$ representation,} characteristic for a cooled electron spectrum (left hemisphere of the left map on Fig.~\ref{fig:indRSG}). The spectral index is roughly \mpo{constant} except at $1$~GeV, where we see a signature of the cooling break. Indeed, electrons close to the shock \mpo{have not had the time to cool at the energies relevant} for the $1$-GeV emission. The formation of two distinct electron populations can be very well seen after $900$ years, in the right hemisphere of the left map of Fig.~\ref{fig:indRSG}, particularly at $1$ and $10$~GeV. The "old" cooled particles are \mpo{far behind the accelerated forward shock, and the  "new" particles close to it} exhibit a much harder spectrum. The difference in spectral index at $10$~GeV reaches unity \mpo{over a distance} of $0.2R_\mathrm{sh}$, which may well be detectable with, e.g., Fermi-LAT, \mpo{although spatial resolution can be an issue}. This would not be a problem for CTA \mpo{measuring at $100$~GeV, but there} the spectral index is again spatially constant, suggesting that we already observe "new" particles injected after the transition. Later on, in the right pie of Fig.~\ref{fig:indRSG}, one can see the "new" population of particles gradually taking over. The spatial variations disappear except for a slight (not-detectable) change in the ring related to the second CD.}

\is{For the WR case (Fig.~\ref{fig:indWR}), the \mpo{GeV spectrum is not immediately affected by} the transition. Right before the transition, after $300$ years, and right after it, at $500$ years, the spectral index is roughly constant in the GeV band. A spatial variation of the spectral index \mpo{emerges when the "old" low-energy component is sufficiently cooled for the cooling signature to appear at photon energies around} a GeV. After $1000$ years (left hemisphere of the right map on Fig.~\ref{fig:indWR}), the $10$-GeV spectral map exhibits a very similar variation as in the immediate post-transition phase of the RSG model. The $1$~GeV map, at the same time, still indicates a featureless uncooled spectrum. The strongest evidence of the formation of two components can be seen at $100$~GeV, \mpo{where the spectral index jumps by $\Delta\alpha\simeq 0.3$ near $0.85R_\mathrm{sh}$}. After $2000$ years the situation changes again: the particles \mpo{have had time to cool, and the characteristic signature shifts to lower energies. The $1$~GeV map at $2000$ years resembles that at $10$~GeV and $1000$ years, and likewise for the $10$-GeV map and the $100$-GeV map. After $2000$ years the} $100$-GeV map is already dominated by the "new" component.}

\section{Summary}
\label{summary}

\is{\mpo{The remnants of core-collapse Supernovae} expand into the circumstellar medium shaped by the wind of their progenitor stars. \mpo{The gas and magnetic-field distributions in the wind bubbles are complex,} which strongly affects not only the evolution of the SNR but also the non-thermal emission produced by relativistic particles accelerated at the SNR shock. We studied the impact of the circumstellar medium and in particular the circumstellar magnetic field on the non-thermal emission of relativistic electrons using the RATPaC software, that is designed for the time- and spatially dependent simulations of particle acceleration at SNR shocks. \mpo{We separately simulate the SNRs of RSG and of WR progenitors.}}

\is{Our simulations demonstrate that during the early evolution of the SNR the electron spectra, and hence those of the non-thermal radiation, might be effectively modified by severe synchrotron losses in the strong magnetic field in the inner wind zone. Synchrotron cooling becomes less efficient with time as the magnetic field declines as $1/r$. Later on, the transition from the \IS{FW} to the \IS{SW}, with the associated abrupt changes in the magnetic-field strength and the gas density, may strongly modify the electron spectrum, which is subsequently reflected in the synchrotron and inverse Compton spectra. In particular, we see the formation of two components of particles yielding a concave spectrum, \mpo{one accelerated in the free wind and the other one produced in the \IS{SW}}. The concavity of the spectrum changes with time, \mpo{and so does the spectral hardness} in the GeV-TeV range. At some times the simulated gamma-ray spectrum produced by electrons resemble that expected from hadronic interactions.}

\is{The two particle populations formed because of the transition to the shocked medium are also spatially separated. This can result in \mpo{a complex and energy-dependent morphology} of the SNR featuring, e.g., multiple bright rings of non-thermal emission. \mpo{An associated} spatial variation of the spectral index can potentially be detected, subject to the angular resolution of the instrument.} \mpon{Additionally, the morphology of IC gamma rays and that of synchrotron X rays do not correlate, as the latter is strongly dependent on the distribution of the magnetic field. It is therefore premature to interpret the absence of such a correlation as an indication for the hadronic origin of the gamma rays.}

\is{\mpo{Variations of the magnetic field influence high-energy electrons through the cooling rate and through the efficiency of confinement at} the shock. 
Additionally, abrupt changes of the gas density, \mpo{e.g. at the wind-termination shock, lead to acceleration or deceleration of the forward shock, depending on the progenitor type, which again has an impact on the maximum attainable} energy. The interaction of the shock with the termination shock provides a reflected shock that can reflect off the CD to later catch up with the forward shock again, triggering an instantaneous acceleration of the latter. \mpo{Events of this type lead to short episodes of strong brightening and/or dimmings in the X-ray and gamma-ray light curves.} Our simulations suggest that such events may be detectable even with current generation instruments.}

\mpo{We used simplified flow profiles for the wind bubbles and in particular concentrate on the first $2000$ years of SNR evolution. We anticipate that similar, but potentially more complex effects can be observed for older SNRs and with refined modelling of the wind bubble \citep[e.g.][]{2020MNRAS.493.3548M}.} {\isref We also emphasize that the actual time scales at which the effects discussed in this work take place strongly depend on the location of the transition point which we assumed to be rather close to the centre of the star.}

\IS{The simulations conducted in this study clearly demonstrate that the properties of the accelerated electrons are shaped by the spatial distribution of the ambient medium and particularly by that of the ambient magnetic field. \mpon{One consequence is a strong temporal variation of the spectra, brightness, and morphology of non-thermal emission} from core-collapse SNRs. On the one hand, it emphasizes the importance of careful analysis of individual SNRs in pursue of the origin of their emission and accordingly the origin of Galactic cosmic rays. On the other hand, it offers an opportunity to obtain valuable insights \mpon{in the properties of SNR from their} non-thermal emission.}

{\isref
\begin{acknowledgments}
This work is based on the research supported in part by the National Research Foundation of South Africa (Grant Number 132276). Robert Brose acknowledges funding from an Irish Research Council Starting Laureate Award (IRCLA/2017/83).
\end{acknowledgments}
}

\bibliography{bibliography}{}
\bibliographystyle{aasjournal}



\end{document}